\documentclass[useAMS,usenatbib,letterpaper]{mn2e}
\usepackage{graphicx}
\usepackage{amsmath}
\usepackage{amssymb}
\usepackage{multirow}

\title[Investigating a model of optimised AGN feedback]{Investigating
  a model of optimised AGN feedback}

\author[E.C.D. Pope] {Edward
  C.D. Pope$^{1}$\thanks{E-mail:ecdpope@uvic.ca}\\$^{1}$Department of
  Physics \& Astronomy, University of Victoria, Victoria, BC, V8P 1A1,
  Canada\\}

\begin{document}

\pagerange{\pageref{firstpage}--\pageref{lastpage} \pubyear{2011}}

 \maketitle

\label{firstpage}

\begin{abstract}
Feedback heating from AGN in massive galaxies and galaxy clusters can
be thought of as a naturally occurring control system which plays a
significant role in regulating both star formation rates and the X-ray
luminosity of the surrounding hot gas. In the simplest case, negative
feedback can be viewed as a system response that is `optimised' to
minimise deviations from equilibrium, such that the system rapidly
evolves towards a steady state. However, a general solution of this
form appears to be incompatible with radio observations which indicate
intermittent AGN outbursts. Here, we explore an energetically
favourable scenario in which feedback is required to both balance
X-ray gas cooling, and minimise the sum of the energy radiated by the
gas and the energy injected by the AGN. This specification is
equivalent to ensuring that AGN heating balances the X-ray gas cooling
with \emph{minimum black hole growth}. It is shown that minimum energy
heating occurs in discrete events, and not at a continuous, constant
level. Furthermore, systems with stronger feedback experience
proportionally more powerful heating events, but correspondingly
smaller duty cycles. Interpreting observations from this perspective
would imply that stronger feedback occurs in less massive objects -
elliptical galaxies, rather than galaxy clusters. One direct
consequence of this effect would be that AGN heating events are
sufficiently powerful to expel hot gas from the gravitational
potential of a galaxy, but not a galaxy cluster, which is consistent
with theoretical explanations for the steepening of the $L_{\rm X}-T$
relation at temperatures below 1-2 keV.
\end{abstract}


\section{Introduction}
The cooling time of X-ray emitting gas in the cores of many massive
galaxies and galaxy clusters is much shorter than the Hubble time. In
the absence of heat sources, the gas will cool and form
stars. However, high-resolution X-ray spectroscopy of galaxies and
clusters has shown that the rate at which gas cools to low
temperatures is significantly reduced compared to preliminary
expectations
\citep[e.g.][]{peterson01,tamura01,xu02,sak,peterson03,kaastra04,peterson06} 
suggesting that the gas is somehow being reheated.

Numerous possible heating mechanisms have been suggested, most notably
energy injection by Active Galactic Nuclei (AGN)
\citep[e.g.][]{bintab,tucker,cfq} and the inward flux of thermal
energy from large radii due to thermal conduction
\citep[e.g.][]{bregdav88,gaetz,zakamska03,pope05}. Quantifying these
heating processes is difficult due our incomplete understanding of the
microphysics of the X-ray emitting plasma \citep[see][for
  example]{cho03,parrish}. Nevertheless, due to its strong temperature
dependence, thermal conduction alone probably cannot provide a general
solution to the cooling flow problem
\citep[e.g.][]{voigt04,pope06,guo}. 

Instead, it is generally assumed that energy input by a central AGN is
predominantly responsible for reheating the gas. This is partly based
on a wealth of observational evidence which indicates that radio AGN
outflows are triggered in response to the thermal state of their
environment \citep[e.g.][]{burns,best05,birzan,dunn05,
  best07,raff08,cav08,mittal}. If true, this suggests that AGN
activity is part of a negative feedback loop which may regulate
properties of its environment.

Theoretical studies have also provided complimentary evidence
highlighting the importance of AGN feedback. For example,
implementations of AGN heating in semi-analytic models of galaxy
formation have shown that, in principle, AGN can both reheat cooling
flows and explain the exponential cutoff at the bright end of the
galaxy luminosity function
\citep[e.g.][]{benson,croton05,bower06,short}. More recently, AGN
heating has been shown to be fundamental in shaping the X-ray
luminosity-temperature of massive galaxies
\citep[e.g.][]{puchwein,bower08,pope09}. 

Broadly speaking, theoretical studies employ AGN heating as an input
with which to control star formation rates and the X-ray luminosity of
the hot gas that permeates massive galaxies and galaxy
clusters. Essentially, this conceptualises AGN feedback as a control
system similar to a thermostat \citep[c.f.][]{stats}. This approach
can be highly informative; AGN feedback is an extremely complex
phenomenon which depends on poorly understood physical processes
occurring across a huge range of spatial scales. As yet, numerical
hydrodynamic simulations do not include sufficient physical processes
to naturally reproduce observations without being complemented by
additional numerical prescriptions. Yet, even if the simulations were
physically complete, the outcomes would still need to be interpreted.

For this reason, we investigate AGN feedback in the most general way,
by expressing its overall effect in terms of well-defined global
quantities, and comparing the main results with observations. For
example, consider the simplest case, in which AGN heating is regulated
only by negative feedback. Negative feedback can be defined as a
response which occurs to oppose its cause. Therefore, unless forced to
do otherwise, it will tend to minimise deviations from some
equilibrium condition. Thus, in general, a system regulated purely by
negative feedback will rapidly tend towards a steady state in which
the controller is permanently active at a constant value. Notably,
this does not appear to be the case for AGN feedback, and it is
imperative to note that minimising deviations from an equilibrium
state is only one measure of the performance of feedback. Depending on
additional constraints it can also be favourable to minimise other
quantities, such as the energy invested in controlling the
system. Importantly, a system regulated by feedback subject to
additional constraints generally behaves quite differently to the
simplest case described above.

The choice of constraint adopted in this article is influenced by two
principle sets of observations. In particular, radio data
\citep[e.g.][]{best05,best08,shab08} can be used to infer the fraction
of time (duty cycle) an AGN spends producing kinetic outflows, which
couple strongly to the ambient gas. Therefore, a feedback model which
yields continual and constant heating would appear to be inconsistent
with data. In addition, any appropriate model must also grow black
holes of the masses we observe. For example, \cite{smbh04} found that,
in some cases, the accretion energy liberated in growing the
supermassive black holes at cluster centres seems to be insufficient
to have offset the ICM X-ray luminosity over the lifetimes of the
clusters. That is, the some black hole masses are lower than
expected. As a result, the aim of this article is to find a plausible
constraint which forces a simple feedback system to exhibit features
that are broadly similar to these observations; heating should proceed
in the form of discrete events and at low black hole growth rates,
while still balancing gas cooling.

Guided by these observations, we investigate an energetically
favourable model in which feedback heating acts to balance gas
cooling, but is also constrained to minimise the total energy output
of the system. This means that feedback minimises the sum of the
energy radiated by the X-ray emitting gas \emph{and} the energy
injected by the AGN, which is equivalent to balancing gas cooling with
the minimum black hole growth. Such constraints are implemented, with
the minimum number of assumptions, by employing optimal control
theory \footnote{Optimal control theory has a wide range of
  applications extending beyond engineering and finance to complex
  biological systems as well as physics. Astrophysically, optimal
  control theory is applied in adaptive optics, but does not appear to
  have been widely adopted in theoretical research, though \cite{hag}
  employed the approach to demonstrate a concise derivation of the
  Tolman-Oppenheimer-Volkoff equation of hydrostatic
  equilibrium.}. Using this approach, we also show that the constraint
prevents continuous, constant AGN power output, but favours discrete
AGN outbursts with a duty cycle which depends only on the feedback
strength. Therefore, despite being largely empirical, the constraint
warrants investigation as a baseline model against which both
observations and numerical simulations can be compared. However, we
note that there may be other equally plausible interpretations, and
explanations, of these observations.

The outline of the article is as follows. Section 2 outlines a simple
feedback model. In Section 3, we investigate the effect of the mimumum
energy constraint which is implemented using optimal control
theory. The findings are discussed in section 4 and summarised in
section 5.

\section{A simple model}

Current interpretation of observations suggests that radio AGN heating
is related to the cooling rate of the hot, X-ray emitting gas
\citep[e.g.][]{best05,birzan,dunn05,nulsen2007}, which itself must
evolve according to the difference between the AGN heating and gas
cooling rates. The precise details of the interaction between a
collimated AGN outflow and the ambient gas may be relevant on
kiloparsec scales, but are much less so on larger scales, since the
injected energy is eventually dissipated more or less isotropically
\citep[see][]{heinz06a}. Therefore, on large spatial scales, AGN
heating can be approximated as supplying thermal energy alone
\citep[e.g.][]{pope09,fabj} and it is reasonable to expect the
heating/cooling system to be explicable by a relatively simple
model. In terms of general expectations, radiative cooling from the
hot gas is a positive feedback process - as energy is radiated, the
gas loses pressure support, contracts and radiates at an accelerating
rate. As a result, the $N$th time derivative, ${\rm d}^{\rm N}L_{\rm
  X}/{\rm d}t^{\rm N}$, must itself be related to $L_{\rm X}(t)$, in
some way. Conversely, heating will cause the X-ray emitting gas to
expand, thereby reducing its luminosity. With this in mind, and
without making any unnecessary assumptions, suppose that $H(t)$ and
$L_{\rm X}(t)$ can be related by a continuous time, linear, $N$-th
order differential equation of the form
\begin{eqnarray}\label{eq:1}
a_{\rm N}\frac{{\rm d}^{\rm N}L_{\rm X}}{{\rm d}t^{\rm N}}+ a_{\rm
  N-1}\frac{{\rm d}^{\rm N-1}L_{\rm X}}{{\rm d}t^{\rm N-1}}
+...\\ \nonumber ...+a_2\frac{{\rm d}^{2}L_{\rm X}}{{\rm d}t^{2}}+
a_1\frac{{\rm d}L_{\rm X}}{{\rm d}t}+a_0 L_{\rm X} = H(t).
\end{eqnarray}
Note that, in its present form, equation (\ref{eq:1}) is a generic
description for the variation of the X-ray luminosity, $L_{\rm X}(t)$,
in response to an externally applied heating rate, $H(t)$. However,
heating rates derived from observations of X-ray cavities inflated by
AGN seem to correlate with the X-ray luminosity of the host cluster
\citep[e.g.][]{birzan,dunn08}, suggesting $H(t) \propto L_{\rm
  X}(t)$. Furthermore, \cite{raff08} \citep[see also][]{cav08} showed
that AGN in cluster centres seem to be activated when the central
cooling time of the ICM is below $\approx 0.5\,{\rm Gyr}$. For this
reason, we can express the heating rate as $H(t) = \alpha(t)k_1 L_{\rm
  X}(t)$, where $k_1$ is a constant of proportionality (the feedback
strength), and $\alpha(t)$ is a parameter which varies between 0 and 1
depending on the central cooling time.

The physical processes that govern how $\alpha(t)$ varies are highly
uncertain, as mentioned in the introduction. They relate to how
material from the ICM reaches the black hole, and how the resulting
energy output from the AGN outflow is dissipated in the surrounding
medium. Rather than trying to solve the gas physics explicitly, we can
pose the problem in a different way. From observations indicating
recurrent discrete AGN outbursts \citep[e.g.][]{best05,best08,shab08},
we conclude that $\alpha(t)$ does vary and is probably close to zero
for a significant fraction of the time, but also close to unity at
other times. This suggests that AGN heating causes the central cooling
time of the ICM to significantly overshoot the critical value of
$\approx 0.5\,{\rm Gyr}$. As a result, it is clear that AGN heating
influences $\alpha(t)$, and $\alpha(t)$ influences AGN heating. The
crucial question then is: how must $\alpha(t)$ be made to vary in
order to explain the observed radio AGN duty cycles
\citep[e.g.][]{best05,best08,shab08}?  One possible solution is
obtained by employing optimal control theory.

For convenience, equation (\ref{eq:1}) can be expressed in matrix form
as a system of first order differential equations. These describe the
evolution of the state variables of the system
\citep[e.g.][]{lincont}, which are defined as $q_1(t)$,
$q_2(t)$....$q_{\rm N}(t)$ where $q_1 = L_{\rm X}$, $q_2(t) = {\rm
  d}L_{\rm X}/{\rm d}t$ up to $q_{\rm N}={\rm d}^{\rm N-1}L_{\rm
  X}/{\rm d}t^{\rm N-1}$. Therefore, the time derivatives of the state
variables are $\dot{q}_1 = q_2$, $\dot{q}_2 = q_3$ up to $\dot{q}_{\rm
  N} = -[a_{\rm N-1}q_{\rm N}+a_{\rm N-2}q_{\rm N-1}+a_{\rm 0}q_{1} -
  H(t)]/a_{\rm N}$. The system can then be written compactly as
\begin{eqnarray}\label{eq:2}
\mathbf{\dot{q}}(t) = \mathbf{A} \mathbf{q}(t) + \mathbf{B} H(t),
\nonumber \\ L_{\rm X}(t) = \mathbf{C} \mathbf{q}(t),
\end{eqnarray}
where $\mathbf{q}(t)$ is the column vector containing the state
variables, $\mathbf{\dot{q}}(t)$ is its time derivative and
$\mathbf{A}$ is called the system matrix. For a system that has $m$
inputs and $p$ outputs, and $N$ state variables, $\mathbf{A}$ is an $N
\times N$ matrix, $\mathbf{B}$ is an $N \times m$ matrix and
$\mathbf{C}$ an $p \times N$ matrix. Therefore, for the case described
above $\mathbf{B}$ is a column vector and $\mathbf{C}$ is a row
vector.

The most convenient way to express the feedback equation is to assume
the control is linearly proportional to all of the states of the
system: $\mathbf{H}(t) = \alpha(t)\mathbf{K} \mathbf{q}(t)$, where
$\mathbf{K}$ is the feedback matrix - a row vector governing the
strength of the feedback \citep[e.g. ][]{feedback}. For a first order
system $\mathbf{K} = [k_1]$, while for a second-order system
$\mathbf{K} = [k_1, k_2]$, where $k_{1}$ and $k_{2}$ denote how the
heating is coupled to $L_{\rm X}$ and $\dot{L}_{\rm X}$,
respectively. To maintain generality, $\mathbf{H}(t)$ has been
expressed as a column vector so that it can, in principle, be related
to any of the derivatives of $L_{\rm X}(t)$ as well as $L_{\rm X}(t)$
itself. However, it seems likely that only the primary coupling
constant, $k_1$, will be non-zero; as a result $H(t)$ is only related
to $L_{\rm X}(t)$.

By assuming feedback control, equation (\ref{eq:2}) reduces to
\begin{equation}\label{eq:3}
\mathbf{\dot{q}}(t) = [\mathbf{A}+\alpha(t)\mathbf{BK}]
\mathbf{q}(t)
\end{equation}
and is said to be controllable if the matrix $\mathbf{K}$ exists such
that it can yield any desired set of roots for the characteristic
polynomial of the closed-loop system, $|\lambda \mathbf{I} -
[\mathbf{A}+\alpha(t)\mathbf{BK}]| = 0$ \citep[e.g.][]{feedback}.

\section{AGN feedback models with additional constraints}

The aim of optimal control theory is to determine how the control
variable must be varied in order to minimise the `cost' associated
with the constraint. The solutions are derived using the calculus of
variations and are, therefore, optimal in the same sense as a solution
of the Euler-Lagrange equations in classical mechanics; by this
definition a natural system must reside in an `optimal' state. As a
result, the behaviour of a system, if interpreted correctly, provides
valuable information about important constraints and the underlying
physical processes at work. In the present example, the control
variable is $\alpha(t)$. Given that $0 \le \alpha(t) \le 1$, we
investigate how it should be varied to minimise the total energy
output of the system. 

As described in the introduction, the motivation for this choice of
constraint is influenced by i) radio data
\citep[e.g.][]{best05,best08,shab08} indicating that AGN heating may
occur in discrete outbursts; ii) mass estimates of black holes in some
Brightest Cluster Galaxies \citep[e.g.][]{smbh04}, which may be
interpreted as having liberated insufficient accretion energy to have
offset the ICM X-ray luminosity over the lifetime of the host
cluster. Below, we outline a model that exhibits features which are
broadly similar to standard interpretations of these observations:
\emph{AGN feedback which acts to minimise the total energy output from
  the system (i.e. gas cooling + AGN heating) will balance gas cooling
  in the form of discrete heating events, and in a way that minimises
  black hole growth}.

\subsection{Optimised AGN feedback}

In this section, we use optimal control theory to determine how
$\alpha(t)$ should be varied in order to minimise the total energy
output of the system. The method is somewhat mathematical, but the
results are very general and there is no need to make any assumptions
about the particular mode of accretion, only that AGN heating is
optimal with respect to the imposed constraint.

In optimal control theory \citep[see][for example]{burghes,optcont},
the control must maximise the so-called objective function, $J$, which
is written as
\begin{equation}\label{eq:obj}
J = \int_{t_0}^{t_1} L[\mathbf{q}(t),\alpha(t)]\,{\rm d}t,
\end{equation}
where, $L$ is the Lagrangian which quantifies the constraints on the
system and is a function of the state and control variables as
shown. For a minimisation, the appropriate objective function is
simply the the maximising objective function multiplied by $-1$.

The Hamiltonian, $F$, of the system is generated by adjoining the
system dynamics, $f[\mathbf{q}(t),\alpha(t)]$, to the Lagrangian, such
that
 \begin{equation}
F = \mathbf{y}(t)f[\mathbf{q}(t),\alpha(t)] +
L[\mathbf{q}(t),\alpha(t)],
\end{equation}
where $\mathbf{y}(t)$ is called the costate vector, and is the dynamic
equivalent of the Lagrange multipliers in static problems of
maximisation. From equation (\ref{eq:3}), the system dynamics can be
written $f[\mathbf{q}(t),\alpha(t)] = \mathbf{\dot{q}}(t) -
[\mathbf{A} + \alpha(t)\mathbf{BK}]\mathbf{q}(t)$.

Optimal control theory also makes use of Pontryagin's maximum
principle, which states that the optimal state $\mathbf{q}^{*}(t)$,
optimal control $\alpha^{*}(t)$ and corresponding costate vector
$\mathbf{y}^{*}(t)$, must maximise the Hamiltonian
\citep[e.g.][]{burghes}. Therefore, the condition for optimality is
\begin{equation}\label{eq:maxp}
\frac{\partial F}{\partial \alpha} = 0.
\end{equation}
For a complete solution, it is also necessary to solve the
Euler-Lagrange equations of the Hamiltonian
\begin{eqnarray}\label{eq:maxp2}
\frac{\partial F}{\partial \mathbf{q}} - \frac{\rm d}{{\rm d}
  t}\bigg(\frac{\partial F}{\partial \mathbf{\dot{q}}}\bigg) = 0
\\ \nonumber \frac{\partial F}{\partial \mathbf{y}} - \frac{\rm
  d}{{\rm d} t}\bigg(\frac{\partial F}{\partial
  \mathbf{\dot{y}}}\bigg) = 0.
\end{eqnarray}
Full solutions can then be obtained by applying the appropriate
boundary conditions, which depend on the specific problem. In this
case, it is assumed that the end time, $t_1$, and its corresponding
state, $L_{\rm X}(t_{1})$, are unknown. This requires an additional
boundary condition, given by $y_{1}(t_1) = 0$, where $y_{1}$ is the
first element of the costate vector. The initial X-ray luminosity is
defined as $L_{\rm X}(t_0)$.

The minimum energy output constraint corresponds to a Lagrangian of
the form $L[\mathbf{q}(t),\alpha(t)] =
-[1+\alpha(t)\mathbf{K}]\mathbf{q}(t)$. Using this, the Hamiltonian
must be
\begin{equation}
F = -[1+\alpha(t)\mathbf{K}]\mathbf{q}(t) +
\mathbf{y}(t)\{\mathbf{\dot{q}}(t) - [\mathbf{A}+\alpha(t)\mathbf{BK}]
\mathbf{q}(t)\}.
\end{equation}
Since $F$ is linear in $\alpha(t)$, it will be minimised when
$\alpha(t)$ takes its largest value (if $\partial F/ \partial \alpha >
0 $), or its smallest possible value (if $\partial F/ \partial \alpha
< 0$) \citep[e.g.][]{optima}.

More specifically, for a first-order system ($N=1$), dimensional
analysis suggests $a_1 \approx -\tau$, where $\tau$ is a
characteristic cooling timescale, with $a_0 \sim 1$. Then, equation
(\ref{eq:maxp}) yields
\begin{equation}\label{eq:ham}
\frac{\partial F}{\partial \alpha} = k_{1}L_{\rm
  X}(t)\bigg[\frac{y_{1}(t)}{\tau}-1\bigg]=0.
\end{equation}
Since the terminal boundary condition for the costate variable is
$y_{1}(t_1) = 0$, equation (\ref{eq:ham}) will be negative late in a
given heating/cooling cycle. Thus, $\alpha(t)$ should optimally be
zero in this limit. Earlier in the cycle, $\lambda(t)$ may be large
enough to make $\partial F/ \partial \alpha > 0 $, in which case the
optimal control is $\alpha(t) = 1$ \citep[e.g.][]{optima}. This is a
highly significant result for all Hamiltonians which are linear in the
control variable: \emph{the optimal control always lies on a boundary
  of the control set, but can switch from one boundary point to the
  other - the so-called `bang-bang' solution.}

Equation (\ref{eq:maxp}) will be satisfied when $y_{1}(t_*) = \tau$,
where $t_{*}$ denotes the time after the start of each cycle at which
$\alpha(t)$ switches from $1 \rightarrow 0$. The value of $t_*$ is
usually obtained by solving equation (\ref{eq:maxp2}), which itself
yields a differential equation for the time-dependence of
$y_{1}$. However, since the initial value of $y_{1}(t_0)$ is unknown,
we must estimate $t_{*}$ by minimising the total energy output with
respect to $t_{*}$, using
\begin{equation}\label{eq:minim}
\frac{\partial}{\partial t_{*}} \int_{t_{0}}^{t_{1}}[L_{\rm X}(t)+
  H(t)]\,{\rm d}t = 0.
\end{equation}
For this calculation, we require the complete solution for the
time-evolution of the X-ray luminosity of the gas, as outlined below.

Since, $\alpha(t)$ is initially 1 and subsequently switches to 0,
there will be two distinct regimes per cycle; the AGN will be active
during the interval $[t_{0}, t_{0} + t_{*}]$, but inactive during the
subsequent interval, $[t_{0} + t_{*}, t_1]$. Then, according to the
boundary conditions, the solution must be
$$ L_{\rm X}(t) = \left\{ \begin{array}{rl} L_{\rm
    X}(t_0)\exp\bigg[(1-k_{1})\frac{(t-t_0)}{\tau}\bigg], & \\ \mbox{
    if $ t_{0} \le t < t_{0}+t_{*}$}\\ \\

L_{\rm
  X}(t_0)\exp\bigg[(1-k_{1})\frac{(t_{*}+t_0)}{\tau}\bigg]\exp\bigg[\frac{[t-(t_0
      + t_{*})]}{\tau}\bigg], & \\ \mbox{ if $ t_{0}+t_{*} \le t <
  t_{1}$}.
	\end{array} \right.
$$ 
Similarly, the AGN heating rate is
$$ H(t) = \left\{ \begin{array}{rl} k_{1} L_{\rm X}(t), & \mbox{ if
    $ t_{0} \le t < t_{0}+t_{*}$}\\ \\

0, & \mbox{ if $ t_{0}+t_{*} \le t < t_{1}$}.
	\end{array} \right.
$$ Then, substituting for $L_{\rm X}(t)$ and $H(t)$ in equation
(\ref{eq:minim}), we find the optimal switching time to be $t_* =
(t_1-t_0) - \tau \ln 2$.

Without making further assumptions, it is not possible to calculate
the precise value of $t_*$. However, in general terms, since AGN
heating probably does not reduce the X-ray luminosity by a large
factor during each cycle, the switching time should scale as $t_* \sim
\tau/(k_{1}-1)$. Indeed, assuming that the AGN is re-triggered when
the ambient conditions return to a state that is similar to the
initial configuration (i.e. $L(t_1) \approx L(t_0)$), a
heating/cooling cycle will have a duration $t_1-t_0 \approx \tau
k_{1}\ln(2)/(k_{1} - 1)$, and switching time, $t_* \approx \tau
\ln(2)/(k_{1} - 1)$. Therefore, for optimal, periodic heating the
duty cycle will be $\delta \equiv t_*/(t_1 - t_0) = 1/k_1$.

Figure 1 shows the time-evolution of the AGN heating rate and X-ray
luminosity for this first-order heating/cooling system under the
assumptions of periodic, optimal feedback. In the figure, the feedback
strength is $k_1 = 10$, so that the duty cycle of AGN heating is
$\delta \approx 0.1$.

\begin{figure*}
\centering
\includegraphics[width=10cm]{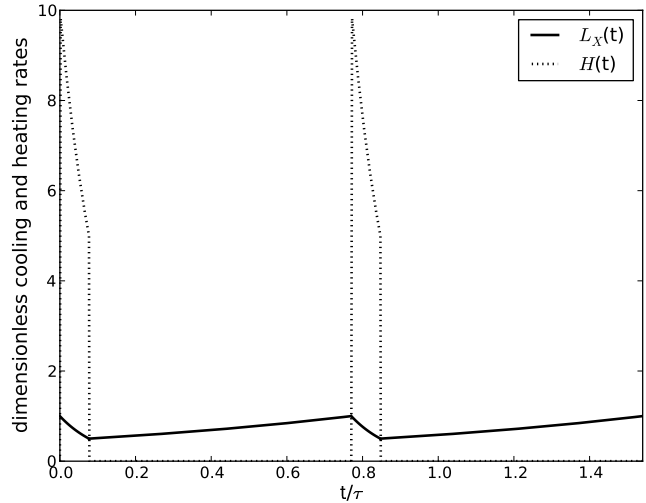}
\caption{Two heating/cooling cycles of the X-ray luminosity, $L_{\rm
    X}(t)$, and AGN heating rate, $H(t)$, for a first-order, linear
  feedback system, subject to the minimum energy constraint. The
  values on the $y$-axis are given by $L_{\rm X}(t)/L_{\rm X}(t_0)$
  and $H(t)/L_{\rm X}(t_0)$, while the x-axis is in units of the
  characteristic timescale of the system, $\tau$. In this example, the
  feedback strength is $k_1 = 10$, so that duty cycle is $\delta
  \approx 0.1$.}
\label{fig:heatcool2}
\end{figure*}

Under the assumption of periodic heating, the energy injected by the
AGN each cycle is the same as the energy radiated by the gas. For
optimal heating, this quantity is $E_{\rm heat} = E_{\rm cool} =
L_{\rm X}(t_0)\tau k_{1}/[2(k_{1}-1)]$. Thus, the total energy
produced each cycle, is $E_{\rm out} = E_{\rm heat} + E_{\rm cool} =
L_{\rm X}(t_0)\tau k_{1}/(k_{1}-1)$. In addition, given that the
time-averaged AGN power output is rate is $\langle H \rangle = E_{\rm
  heat}/(t_1-t_0)$, the time-averaged black hole growth rate must be
$\langle \dot{M}_{\rm BH}\rangle = \langle H \rangle/(\eta c^{2}) =
L_{\rm X}(t_0)/(\eta c^{2} \ln 4)$, where $\eta$ is the standard
accretion efficiency and $c$ is the speed of light. During the
accretion phase, the average accretion rate of the black hole growth
rate is a factor $k_1 = 1/\delta$ greater, and can be approximated by
$\dot{M}_{\rm BH} = L_{\rm X}(t_0)/(\delta \eta c^{2} \ln
4)$. Consequently, systems with characteristically small duty cycles
will experience brief phases of rapid black hole growth. In contrast,
black hole growth in systems with duty cycles closer to unity will be
much more uniform.

\section{Discussion}

Observationally, the radio AGN duty cycle is found to increase steeply
with the stellar mass of the host galaxy \citep[e.g.][]{best05}, and
is higher still for Brightest Cluster Galaxies \citep[][]{best08}. In
the context of the model presented here, this suggests that the
feedback strength decreases with increasing stellar mass.

If the argument above holds, the ratio of the instantaneous heating
and cooling rates ($H/L_{\rm X} = k_1$) will also decrease with
increasing stellar mass of the host galaxy. Consequently, lower mass
systems must experience proportionally more powerful heating
events. As a result, the momentum and energy supplied by the AGN would
be much more capable of expelling X-ray emitting material from the
gravitational potential of an elliptical galaxy than a galaxy
cluster. This scenario would appear to be consistent with recent
theoretical interpretations of the steepening of the $L_{\rm X}-T$
relation below 1-2 keV \citep[e.g.][]{puchwein,bower08,pope09}.

A plausible relationship between $k_1$ and the average temperature of
the ambient gas, is as follows. When activated, the AGN heating rate
is assumed to be proportional to the X-ray luminosity, such that $H(t)
= \eta \dot{M}_{\rm BH}c^{2} = k_1 L_{\rm X}(t)$, where $\dot{M}_{\rm
  BH}$ is the black hole fuelling rate, and the other quantities are
as previously defined. The classical mass flow rate for an X-ray
emitting atmosphere with luminosity $L_{\rm X}(t)$, and average
temperature $T$, is defined as $L_{\rm X} = (5/2)k_{\rm B}T
\dot{M}_{\rm clas}/\mu m_{\rm p}$, where $k_{\rm B}$ is the Boltzmann
constant and $\mu m_{\rm p}$ is the mean mass per particle. Thus, if a
fraction, $\phi$, of the classical mass flow rate reaches the
supermassive black hole, we can write $\dot{M}_{\rm BH} = \phi
\dot{M}_{\rm clas}$, and it can be shown that $k_1 = 2 \phi \eta \mu
m_{\rm p} c^{2}/(5 k_{\rm B}T)$. Therefore, stronger feedback (and
smaller duty cycles) should be observed in cooler, less massive
systems, as is consistent with standard interpretations of radio AGN
observations \citep[][]{best05,best08,shab08}.

To calculate a numerical value for $k_1$, we assume a canonical
accretion efficiency of $\eta = 0.1$, which yields $k_1 = 1/\delta
\approx 260 (\phi/0.01)(T/10^7\,{\rm K})^{-1}$. For low mass
elliptical galaxies, \cite{best05} found $\delta \sim 10^{-4}-10^{-3}$
which, in the context of the present model suggests $k_1 \sim
10^{3}-10^{4}$. Thus, for a characteristic X-ray temperature of $T
\sim 10^6\,{\rm K}$, a value of $k_1 \sim 10^{3}-10^{4} $ can only be
obtained if $\phi \sim 10^{-2}$. In contrast, the duty cycle of radio
AGN activity in Brightest Cluster Galaxies is $\delta \sim 0.1-1$
\citep[e.g.][]{best08}, suggesting that $k_1 \sim 1-10$. Given that
the typical X-ray temperature in such systems is $T \sim 10^8\,{\rm
  K}$, the fraction of the classical cooling flow rate accreted by the
black hole must be $\phi \sim 10^{-4}-10^{-3}$.

If the reasoning outlined above approximates reality, we might infer
that the accreted fraction, $\phi$, tends to decrease in hotter (more
spatially extended) systems. The simplest qualitative explanation for
such a phenomenon is that the black hole predominantly accretes
material from the central region of a galaxy or cluster. Therefore,
the fraction of material accreted from the entire cooling flow will be
governed by the size of this central region relative to the size of
the region within which the classical cooling flow rate applies.

\section{Summary}

The aim of this article has been to demonstrate how standard
techniques optimal control theory can be used to investigate some of
the fundamental characteristics of feedback which may be applicable to
AGN heating in massive galaxies and galaxy clusters. Although it has
not yet been possible to construct a completely satisfactory model for
AGN heating, the application of optimal control theory has elucidated
some potentially important features of constrained feedback
systems. The main findings are described below.

\begin{enumerate}
\item Optimal control theory provides a way to impose additional
  constraints on the system, with the minimum number of
  assumptions. In general terms, constraints are extremely useful for
  conceptualising the interaction between the AGN and its environment
  and interpreting observational signatures. More specifically, we
  have shown that AGN heating (regulated by feedback) which acts to
  minimise the total energy output of the system, occurs in the form
  of discrete, periodic events. This constraint ensures that AGN
  heating \emph{balances gas cooling with the minimum black hole
    growth} and produces behaviour which is broadly compatible with a
  range of observations. However, this is only one interpretation and
  explanation of the observations. In principle, there may be other
  effects or constraints which achieve similar outcomes.

\item The duty cycle for periodic AGN activity in a first-order,
  linear feedback system is $\delta \approx 1/k_1$, where $k_1$ is the
  feedback strength. If this interpretation is applied to observations
  which show radio AGN duty cycles increasing with the stellar mass of
  the host galaxy \citep[][]{best05,best08,shab08}, we would infer
  that the feedback strength decreases as the system stellar mass
  increases. In turn, this would mean that AGN outbursts in lower mass
  systems are proportionally stronger than those in more massive
  environments so that AGN heating events may be sufficiently powerful
  to expel hot gas from the gravitational potential of a galaxy, but
  not a galaxy cluster. A description of this sort is in qualitative
  agreement with recent theoretical explanations for the steepening of
  the $L_{\rm X}-T$ relation for temperatures below 1-2 keV
  \citep[e.g.][]{puchwein,bower08,pope09}.
\end{enumerate}

\section{Acknowledgements}
The author would like to thank CITA for funding through a National
Fellowship.

\bibliography{database} \bibliographystyle{mn2e}

\label{lastpage}
 \end{document}